\documentclass[fleqn,10pt]{wlscirep}

\usepackage{graphicx}
\usepackage{cite}
\usepackage{color}
\usepackage{amsmath}
\usepackage{here}
\usepackage{epstopdf}

\newcommand{\be}{\begin{equation}}
\newcommand{\ee}{\end{equation}}
\newcommand{\bea}{\begin{eqnarray}}
\newcommand{\eea}{\end{eqnarray}}

\newcommand\pictc[5]{\begin{figure}
                       \centerline{\vspace{-1mm}
\includegraphics[width=#1\columnwidth,height=0.7\textheight,keepaspectratio]{#3}}
                       \protect\caption{\protect\label{#4} #5}\vspace{-3mm}
                    \end{figure}            }

\newcommand\pict[4][1]{\pictc{#1}{!tb}{#2}{#3}{#4}}
\newcommand\rpict[1]{Fig.\ref{#1}}

\newcounter{Fig}

\title{Geometric interpretations for resonances of plasmonic nanoparticles}

\author[1,2,3,*]{Wei Liu}
\author[3]{Rupert F. Oulton}
\author[2]{Yuri S. Kivshar}
\affil[1]{College of Optoelectronic Science and Engineering, National University of Defense
Technology, Changsha, Hunan 410073, China}
\affil[2]{Nonlinear Physics Centre, Centre for Ultrahigh-bandwidth Devices for Optical Systems (CUDOS), Research School of Physics and Engineering, Australian National
University, Canberra, ACT 0200, Australia}
\affil[3]{The Blackett Laboratory, Department of Physics, Imperial College London, London SW7 2AZ, United Kingdom}
\affil[*]{Corresponding author: wei.liu.pku@gmail.com}

\begin{abstract}
The field of plasmonics can be roughly categorized into two branches:  surface plasmon polaritons (SPPs) propagating in waveguides and localized surface plasmons (LSPs) supported by scattering particles. Investigations along these two directions usually employ different approaches, resulting in more or less a dogma that the two branches progress almost independently of each other, with few interactions. Here in this work we interpret LSPs from a Bohr model based geometric perspective relying on SPPs, thus establishing a connection between these two sub-fields. Besides the clear explanations of conventional scattering features of plasmonic nanoparticles, based on this geometric model we further demonstrate other anomalous scattering features (higher order modes supported at lower frequencies, and blueshift of the resonance with increasing particle sizes) and multiple electric resonances of the same order supported at different frequencies, which have been revealed to originate from backward SPP modes and multiple dispersion bands supported in the corresponding plasmonic waveguides, respectively. Inspired by this geometric model, it is also shown that, through solely geometric tuning, the absorption of each LSP resonance can be maximized to reach the single channel absorption limit, provided that the scattering and absorption rates are tuned to be equal.
\end{abstract}
\begin{document}

\flushbottom
\maketitle
%
%
\thispagestyle{empty}


\section*{Introduction}

Fueled by the observation of extraordinary transmission through thin metal films~\cite{Ebbesen1998_Nature} and the rapid expansion of the field of metamaterials~\cite{Zheludev2012_NM}, the old subject of plasmonics has gained a strong renewed impetus and has been experiencing an unprecedented explosive growth~\cite{Zayats2005,Atwater2007_SA,SAM,Gramotnev2010}. Great progress has been made not only in the branch of propagating SPPs, such as plasmonic circuitry~\cite{Gramotnev2010}, plasmonic nanolaser~\cite{Oulton2009}, plasmonic beam shaping~\cite{Liu2010,Huidobro2010,Liu2011_PRB} and so on, but also in the other branch of LSPs, \textit{e.g.} superscattering~\cite{Ruan2010_PRL,Ruan2011_APL}, efficient scattering pattern shaping~\cite{Liu2014_CPB,Liu2012_ACSNANO,Liu2013_OL2621} and even LSPs based biological and medical applications~\cite{Atwater2007_SA,Huschka2011_JACS}, to name but a few. However, due to the contrastingly different characteristics (propagating or localized) of the excited states of SPPs and LSPs, and  that investigations conducted in these two sub-fields usually employ quite different approaches and techniques, usually SPPs and LSPs are discussed separately~\cite{Zayats2005,SAM}, which consequently leads to more or less a dogma that the two branches develop almost independently, with few interactions between each other.

 Recently there have been some demonstrations relying on the interactions of these two sub-fields, such as the achievements in superscattering~\cite{Ruan2010_PRL,Ruan2011_APL} and transformation optics~\cite{Pendry2012_Science}, nevertheless general and comprehensive studies on the links between them are still not available. Such investigations have been further necessitated by the need of more intuitive understanding of the LSP resonances supported by various plasmonic particles. It is well known that the topic of particle scattering plays a fundamental role in countless related applications~\cite{Kerker1969_book,Zayats2005,SAM,Huschka2011_JACS}, however an intuitive understanding of LSPs is not so direct, even for the simplest structures of homogeneous spheres and cylinders. Although Mie theory can give all the information required for the descriptions of LSP resonances of basic spherical or cylindrical plasmonic nanoparticles, the basic physical mechanism is somehow shadowed by the kind of complicated formulas and thus an intuitive and clear picture is still not accessible.  Moreover, the topic of light scattering by plasmonic particles is merging rapidly with the current booming fields of graphene~\cite{Geim2007_NM} and topological states of light and matter~\cite{Yin2013_science,Khanikaev2013_NM}, making the searching for an intuitive geometric physical model even more urgent.

 Here in this paper we provide an intuitive geometric picture for LSPs based on the Bohr model,  which requires an integral number of $2\pi$ phase accumulation along an enclosed propagating loop to support a well defined localized resonance~\cite{Landau1977_book,Yang2012_PT}. The propagating SPPs play a fundamental role in this geometric model, which can explain simply and clearly many conventional scattering features of plasmonic nanoparticles, including well defined LSP resonance in the subwavelength regime, higher (lower) order modes supported at higher (lower) frequencies, and the redshift (blueshift) of the resonances with increasing (decreasing) particle sizes. Based on this geometric model, we further demonstrate many other anomalous scattering features of plasmonic nanoparticles: higher (lower) order modes supported at lower (higher) frequencies and the blueshift (redshift) of the resonances with increasing (decreasing) particle sizes. Those anomalous scattering features have been revealed to originate from the backward SPP modes supported by the corresponding plasmonic waveguides. At the same time, we also show the existence of multiple electric resonances of the same order supported at different frequencies, which have been proved to originate from the multiple dispersion bands of the corresponding waveguides. At the end to further exemplify the effectiveness of the geometric model, we demonstrate that at a fixed resonance frequency, the absorption of a LSP resonance can be maximized to reach the single channel absorption limit through pure geometric tuning, when the scattering and absorption rates are tuned to be equal.
\section*{Results}

%
%
%
%
%

\subsection*{Geometric model for Localized surface plasmons}

 \subsubsection*{The application of Bohr condition to localized surface plasmons}

 Basically LSPs are localized resonances and are characterized by a set of discrete resonant frequencies~\cite{Kerker1969_book,Zayats2005,SAM}. Similar to most localized resonances, the existence of the resonance requires the satisfaction of the Bohr condition~\cite{Landau1977_book,Yang2012_PT}. According to the Bohr model, to support a well defined resonance, the length of the enclosed orbit should contain an integral number of the de Broglie wavelengths. It was exactly through Bohr condition that the quantum states of Hydrogen atom were geometrized. The Bohr model can be applied in optics and the Bohr condition can be expressed as:
\begin{equation}
\label{bohr condition}
\oint {n(\textbf{r})} k_0d\textbf{r} = 2m\pi, 
\end{equation}
where $k_0$ is the angular wave-number in free space, $n(\textbf{r})$ is position dependent effective refractive index, and m is an integer, which corresponds to the order of the modes supported ($m=1,~2,~3,~...$ corresponds to dipole, quadrupole, hexapole...). This condition basically means that a resonance requires that the phase accumulation along an enclosed optical path should be an integral number of $2\pi$. One challenge for the application of this model is to decide the loop along which  the phase is accumulated. For example, related analysis has been performed for the whispering-gallery modes of dielectric spheres~\cite{Roll2000_JOSAA}. However, the process to decide the phase accumulation loop is so complicated, especially for the lower order modes, rendering the calculations as complicated as those of rigourous Mie theory and thus losing the simplicity of this model.

In contrast, SPP modes in plasmonic waveguides are propagating coupled states of electrons and photons, with a well defined wave-vector along the propagation direction and with most of the energy confined at the metal-dielectric boundary~\cite{Zayats2005,SAM,Gramotnev2010}. Such features of SPP modes make it easier to decide both the effective wave-number and the phase accumulation path~\cite{Beck2011_OE_Resonant,Hasan2011_relating}, and as a result Eq. (\ref{bohr condition}) can then be simplified as:

\begin{equation}
\label{lsp_bohr condition}
\oint {{k_{\rm spp}}d\textbf{r}}  = 2m\pi,
\end{equation}
where $k_{\rm spp}$ is the angular wave-number for the SPPs in a corresponding plasmonic waveguide. Similar analysis, such as the \textit{Wentzel}-\textit{Kramers}-\textit{Brillouin} approximation~\cite{Marcuse1972}, have been performed for plasmonic waveguides in the transverse direction, and explains intuitively the origins of propagating plasmonic modes of different orders excited on the surface of plasmonic nanowires~\cite{Schmidt2008_OE,Catrysse2009_APL}.

\pict[0.5]{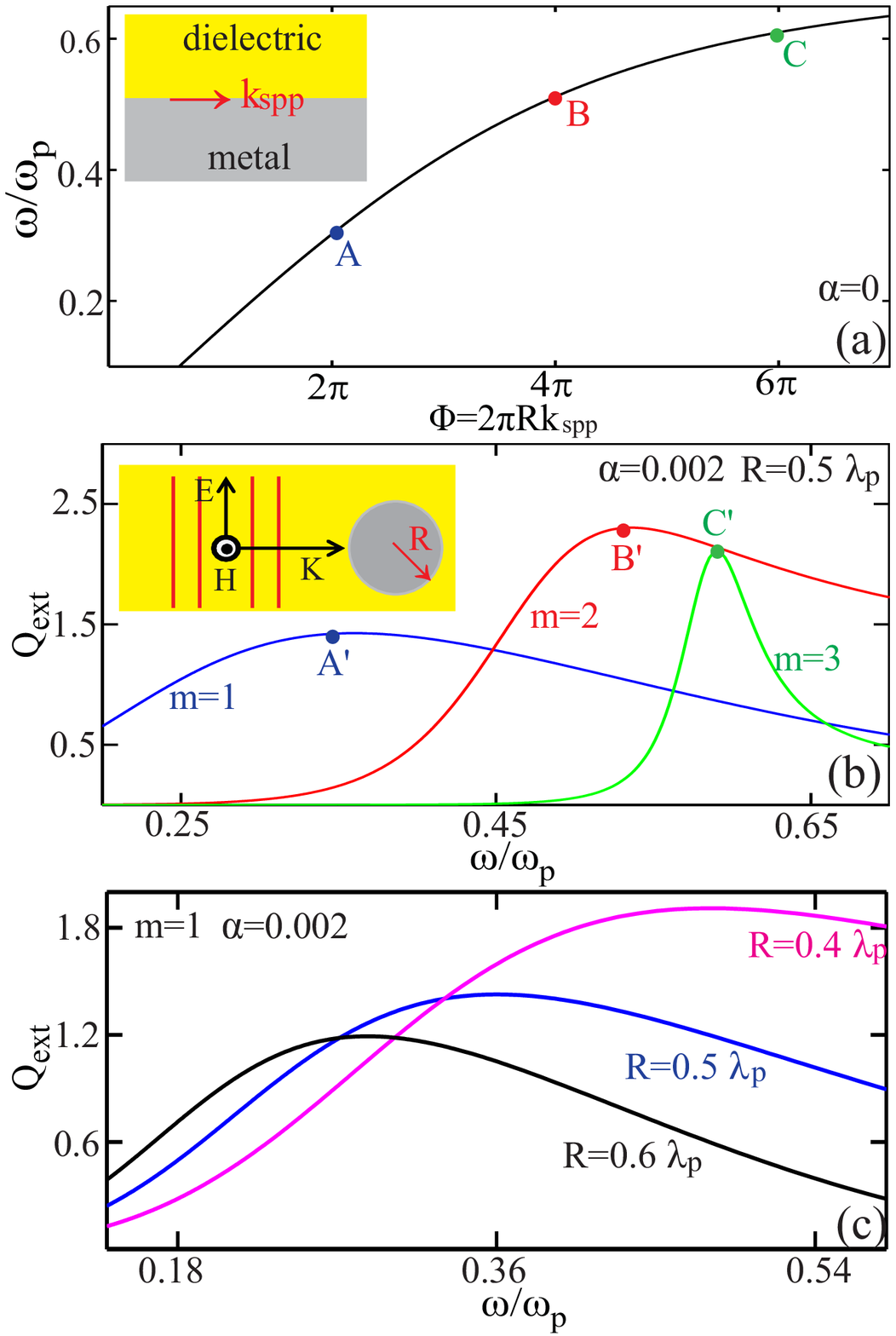}{fig1}{\small (a) Dispersion of the SPP modes supported by the semi-infinite metal-dielectric structure (inset; lossless Drude model is employed to characterize the metal: $\alpha=0$). At points A, B, C the Bohr condition is satisfied with $m=1,2,3$, at $\omega/\omega_p=0.303, ~0.513$ and $0.61$  respectively. (b) Extinction efficiency spectra of the metal nanowire of radius $R=0.5\lambda_p$, showing the contributions from the first three electric modes, which resonate at $\omega/\omega_p=0.356, ~0.536$ and $0.591$  respectively (the central resonant positions are denoted by points A', B' and C' respectively). The incident plane wave is polarized in plane and the lossy Drude model is employed for the metal [inset, $\alpha=0.002$, as is the case in (c)]. (c) Extinction efficiency spectra for the electric dipole modes ($m=1$) supported by three nanowires of different radii: $R=0.4\lambda_p,~0.5\lambda_p,~0.6\lambda_p$.}

 \subsubsection*{Normal dispersions and conventional scattering features}

 To clearly explain the links between SPPs and LSPs, and to verify the validity of Eq. (\ref{lsp_bohr condition}) for LSPs, we start with the simplest case in \rpict{fig1}: single-layered metal nanowire [inset of \rpict{fig1}(b)] and its corresponding plasmonic waveguide of semi-infinite metal-dielectric structure [inset of \rpict{fig1}(a)]. To perform proof-of-concept demonstrations of the model and at the same time not necessarily lose the generality, throughout this work, we characterize the metal with the Drude model:  $\varepsilon_{\rm{m}} = 1 - {\omega_p^2}/(\omega^2
+ i\omega \omega_c)$, where ${\omega}$ is the angular frequency of light, ${\omega_p}$ is the plasma frequency (with corresponding plasmon wavelength of $\lambda_p$), ${\omega_c }$ is the
collision frequency and the loss factor is defined as $\alpha=\omega_c /\omega_p$.  The dispersion relation for the SPP mode of the semi-infinite metal-dielectric structure [inset of \rpict{fig1}(a)] is: ${k_{{\rm{spp}}}} = {k_0}\sqrt {{{{\varepsilon _m}{\varepsilon _d}} \over {{\varepsilon _m} + {\varepsilon _d}}}}$, where $\varepsilon _m$ and $\varepsilon _d$ are the permittivities of the metal and dielectric respectively, and $k_{\rm spp}$ is the effective wavenumber of the SPP mode supported. In \rpict{fig1}(a) we show the dispersion curve ($\varepsilon_{\rm{d}}=1$), where the horizontal axis of angular wave-number is replaced by the phase accumulation along the circumference of the corresponding nanowire in \rpict{fig1}(b): $\Phi  = 2\pi R{k_{{\rm{spp}}}}$. The resonant positions of the first three electric modes predicted by the Bohr model are indicated by points A, B and C, which corresponds to $m=1,2, 3$ respectively in Eq. (\ref{lsp_bohr condition}). For comparison, \rpict{fig1}(b) shows the extinction efficiency spectra of the corresponding nanowire ($R=0.5\lambda_p$) with incident plane wave polarized in plane, to guarantee the excitation of LSP resonances. We note that for simplicity, throughout this work for the dispersion calculation we adopt the lossless Drude model ($\alpha=0$), and for the scattering calculation we adopt the lossy Drude model ($\alpha=0.002$). The positions of extinction peaks for the first three electric modes in \rpict{fig1}(b) (marked by by points A', B' and C' respectively) agree well with the points pinpointed in \rpict{fig1}(a) (points A, B, C, respectively), verifying the validity of the Bohr model.

For dielectric particles, a well defined resonance can be supported only when the size of the particle is comparable with the effective wavelength of incident waves, in order to provide long enough optical path for sufficient phase accumulation~\cite{Kerker1969_book}, as is indicated by Eq. (\ref{bohr condition}). In sharp contrast to dielectric particles, even deep subwavelength plasmonic nanoparticles are able to support well defined resonances~\cite{Kerker1969_book,Zayats2005,SAM}. According to the geometric picture described above, this is due to the fact that the SPP mode can have much larger effective wavenumber and thus sufficient phase can be accumulated over much shorter optical paths.  Other well known conventional scattering features of plasmonic nanoparticles are quite similar to those of dielectric particles, including that: (i) higher (lower) order modes are supported at higher (lower) frequencies [as shown in \rpict{fig1}(b)]; (ii) the resonances are redshifted (blueshifted) with increasing (decreasing) particle sizes. We exemplify the latter point in \rpict{fig1}(c), where the extinction efficiency spectra of the electric dipole modes ($m=1$) are shown for three nanowires of different radii $R=0.4\lambda_p,~0.5\lambda_p,~0.6\lambda_p$. According to Eq. (\ref{lsp_bohr condition}) it is easy to figure out that all those conventional scattering features originate from the normal dispersion of the SPP modes ($d\omega/dk_{\rm spp} > 0$) in the plasmonic waveguides as shown in \rpict{fig1}(a).

\subsubsection*{Backward modes and anomalous scattering features}

\pict[0.5]{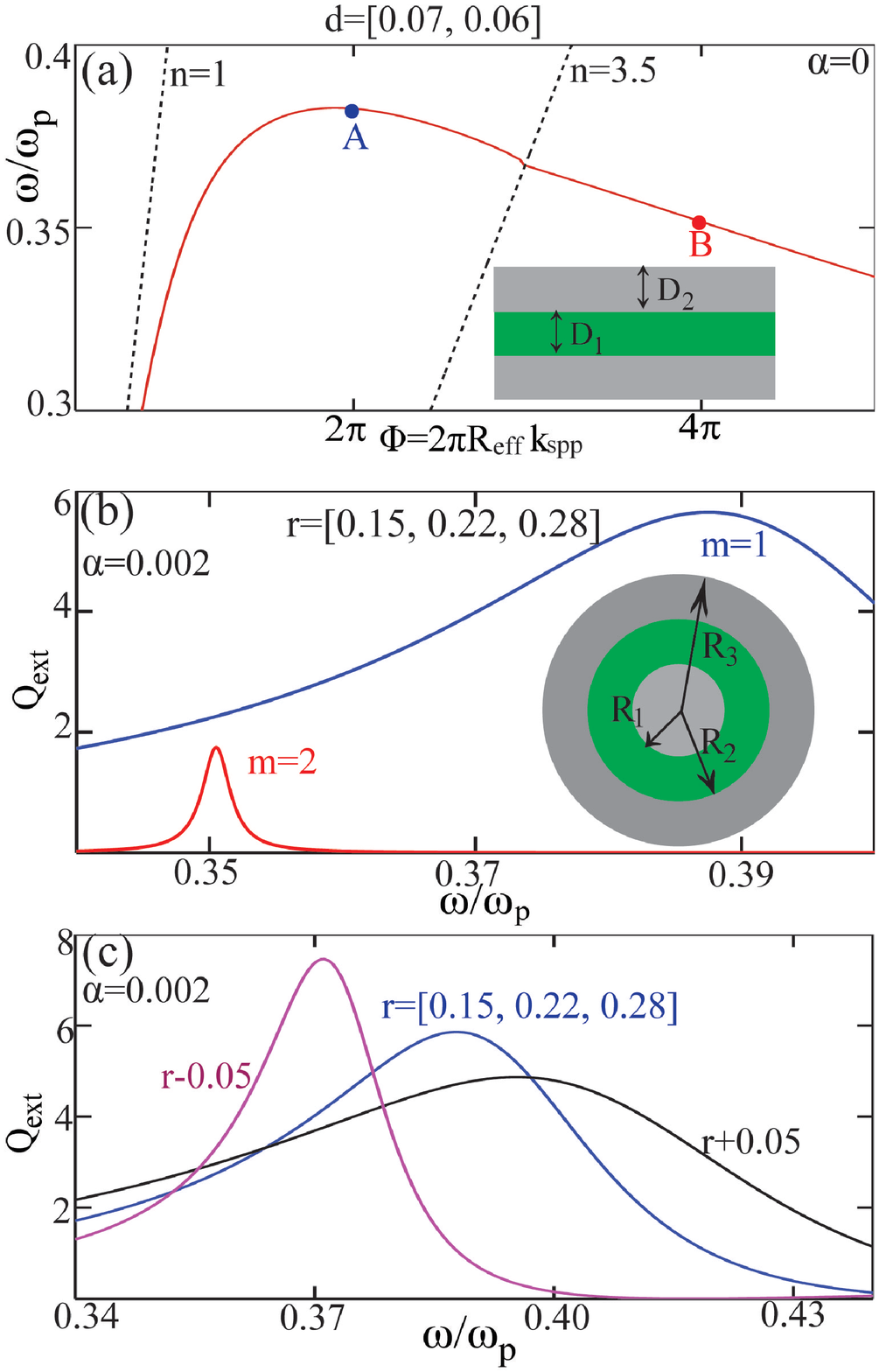}{fig2}{\small (a) Dispersion of the backward SPP modes in the three-layered metal-dielectric planer waveguide (inset). The size parameter is $d=[0.07,~0.06]$ and at points A, B the Bohr condition is satisfied with $m=1, 2$, at $\omega/\omega_p=0.382$ and $0.352$ respectively. The dashed black lines denote the light-lines of $n=1$ and $n=3.5$. Grey and green layers of the inset denote metal and dielectric ($n=3.5$) layers respectively. (b) Extinction efficiency spectra of the corresponding three-layered nanowire of size parameter $r=[0.15,~0.22,~0.28]$, showing the contributions from the first two electric modes, which resonate at $\omega/\omega_p=0.388$ and $0.351$ respectively. The scattering configuration is the same as that shown in the inset of \rpict{fig1}(b). (c) Extinction efficiency spectra for the electric dipole modes ($m=1$) supported by three-layered nanowires of different size parameters: $r=[0.15,~0.22,~0.28]$, $r-0.05$ and $r+0.05$.}

It is well known that the dispersions of SPP modes in plasmonic waveguides can be easily engineered through geometric tuning~\cite{Zayats2005,SAM,Shin2006_PRL,Lezec2007_science,Dionne2008_OE,Archambault2012_PRL}. Besides the normal dispersion, both flat band ($d\omega/dk_{\rm spp} = 0$) and anomalous dispersion (backward modes, $d\omega/dk_{\rm spp} < 0$) can be obtained. According to Eq. (\ref{lsp_bohr condition}), unconventional scattering features can be obtained through dispersion engineering of the modes within the corresponding plasmonic waveguides. For example, for a flat band of SPP mode, it is quite natural to expect that  modes of different orders can be overlapped at the same frequency and the resonant frequency is fixed and independent of the particle size. Such unusual scattering features have already been observed, with a noticeable example of the recent demonstration of superscattering~\cite{Ruan2010_PRL,Ruan2011_APL}.

Beyond the flat-band dispersion of SPP modes, even backward modes can be obtained in plasmonic structures when the direction of overall energy flow is contradirectional to that of the wave-vector~\cite{Shin2006_PRL,Lezec2007_science,Dionne2008_OE}. The inset of \rpict{fig2}(a) shows the three-layered plasmonic waveguide we investigate. The green layer denotes dielectric layer, here for which we assign refractive index of $n=3.5$ (\textit{e.g.} GaAs, Si or Ge). The bottom metal layer is supposed to be semi-infinite and the background is filled with air of $n=1$. The corresponding scattering nanowire of this three-layered waveguides is shown in the inset of \rpict{fig2}(b) [scattering configuration is the same as that in \rpict{fig1}(b)]  and the size parameters satisfy: $D_1=R_2-R_1$ and $D_2=R_3-R_2$. Here we define normalized size parameters for both the waveguide and the nanowire: $d=[D_1,~D_2]/\lambda_p$, $r=[R_1,~R_2,~R_3]/\lambda_p$. For the waveguide, as there are two metal-dielectric boundaries, we approximate the effective length of the phase accumulation path as $2\pi R_{\rm eff}=\pi(R_1+R_2)$. The specific size parameter for the waveguide is $d=[0.07,~0.06]$ and for the nanowire is $r=[0.15,~0.22,~0.28]$. The dispersion of backward SPP mode (TM modes) supported in the waveguide is shown in \rpict{fig2}(a), with two points A and B pinpointed, where the Bohr condition is satisfied with $m=1, 2$ respectively. The extinction efficiency spectra of the corresponding nanowire is shown in \rpict{fig2}(b), with contributions from the first two electric modes shown. As expected, the positions of the spectra peaks agree well with the positions of pinpointed points in \rpict{fig2}(a).  In sharp contrast to the results in \rpict{fig1}(b), here the higher order mode ($m=2$, quadruple mode) is supported at a lower frequency, and lower order mode ($m=1$, dipole mode) is supported at a higher frequency. It is obvious that such anomalous scattering comes from the anomalous dispersion shown in \rpict{fig2}(a), where the backward mode plays the fundamental role.  We further investigate the size dependence of the resonances originating from backward modes, and show the results in \rpict{fig2}(c). The extinction efficiency spectra of the dipole modes ($m=1$) supported by three multi-layered nanowires with different size parameter are shown with: $r=[0.15,~0.22,~0.28]$, $r-0.05$ (r=[0.10,~0.17,~0.23]) and $r+0.05$  (r=[0.20,~0.27,~0.33]). It is clear that the resonances are  blueshifted (redshifted) with increasing (decreasing) particle sizes, which is opposite to the size-scaling trend shown in \rpict{fig1}(c). This anomalous feature also has its origin in the existence of the backward mode. We note here that although similar unusual scattering features mentioned above have been partly observed before~\cite{Peng2010_PNAS,Yang2012_NP450}, but unfortunately the physical mechanism behind had not been clearly revealed or systematically investigated.

\subsubsection*{Multiple dispersion bands induced multiple modes of the same order}

\pict[0.75]{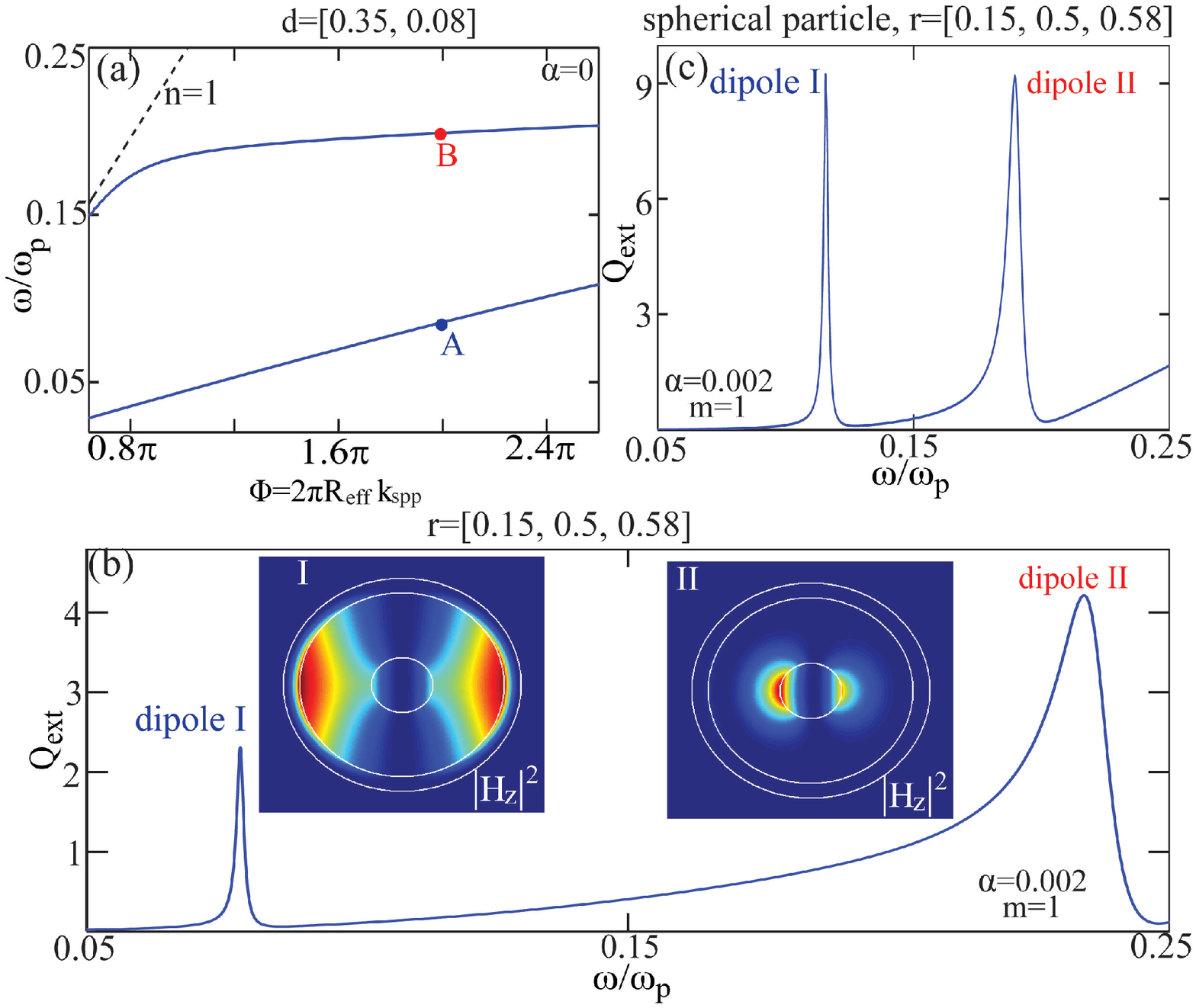}{fig3}{\small (a) Dispersions of SPP modes in the three-layered metal-dielectric structure and two dispersion bands are simultaneously exhibited. The size parameter is $d=[0.35,~0.08]$. At points A, B the Bohr condition is satisfied with the same integer of $m=1$, at $\omega/\omega_p=0.085$ and $0.2$ respectively.  The dashed black line denote the light lines of $n=1$. (b) Extinction efficiency spectra of the two electric dipole modes supported by the corresponding three-layered nanowire of size parameter $r=[0.15,~0.5,~0.58]$. The scattering configuration is the same as that shown in the inset of \rpict{fig1}(b). Two dipoles modes are spectrally centered at $\omega/\omega_p=0.078$ and $0.234$ respectively, with the near-field distributions (out-of-plane magnetic field intensity) shown in insets. (c)  Extinction efficiency spectra of the two electric dipole modes supported by the corresponding three-layered nanosphere of size parameter $r=[0.15,~0.5,~0.58]$ (the definition of the geometric parameters and the scattering configuration is the same as that of the nanowire shown in \rpict{fig2}(b), except that the scattering particle is now a three-dimensional nanosphere).}

Up to now, we have discussed only the LSP resonances based on SPP modes of a single dispersion band. According to Eq. (\ref{lsp_bohr condition}), when several dispersion bands are simultaneously involved, different sets of modes of the same order can be supported. The number of the modes of the same order is decided by the number of the dispersion bands. \rpict{fig3}(a) shows the dispersion curves of the three-layered metal-dielectric waveguide of the size parameter of $d=[0.35,~0.08]$, where two dispersion bands are supported. Points A and B predict the spectral position of the localized modes of the corresponding three-layered nanowire of the the size parameter of $r=[0.15,~0.5,~0.58]$. At both points the Bohr condition is satisfied with the same integer of $m=1$, indicating that two electric dipoles should be supported at different frequencies. To verify the prediction shown in \rpict{fig3}(a), in \rpict{fig3}(b) we show the extinction efficiency spectra of the dipole modes supported by the corresponding nanowire.  As  predicted, it is clear that two dipoles are supported at different frequencies: dipoles \uppercase\expandafter{\romannumeral1} and \uppercase\expandafter{\romannumeral2} correspond to the points A and B in \rpict{fig3}(a) respectively. To further clarify the different origins of the two dipole modes, we show the near-field distributions [out-of-plane magnetic field intensity, insets of \rpict{fig3}(b)] of the two modes at the resonant frequencies of $\omega/\omega_p=0.078$ and $0.234$ respectively. It is clear that though the two modes are of the same order ($m=1$) and are both electric dipole modes, they show quite different near-field distributions. This is due to the fact that the two dipole modes originate from different SPP modes of different dispersion bands [see \rpict{fig3}(a)], which themselves correspond to different near-field distributions within the three-layered waveguides.  Multiple electric modes of the same order play a fundamental role in the recent interpretation of scattering dark states, but unfortunately the physical origin had not been revealed~\cite{Hsu2014_NL}. We emphasize that some other models that deal with the LSP modes of plasmonic nanoparticles, such as the the hybridization model put forward in Ref.\cite{Prodan2003}, can explain the existence of multiple modes of the same order, but in comparison the Bohr model presented in this paper is more direct and simpler. We also note that in \rpict{fig3} we discuss only the dipole modes and this principle can be easily applied to other higher order modes.

Based on the results shown in \rpict{fig3}, we could comment further on our approximation of the effective radius of the phase accumulation path:  $R_{\rm eff}=(R_1+R_2)/2$, which actually assumes that the center of field distribution is in the middle of the two metal-dielectric interfaces.  However, according to the near-field distributions of the two electric dipoles shown in  \rpict{fig3}(b), at points A and B the approximation underestimates and overestimates the effective length of the phase accumulation path respectively. This is because for dipole \uppercase\expandafter{\romannumeral1} (\uppercase\expandafter{\romannumeral2}), most of the fields are confined closer to the outer (inner) metal-dielectric interface. Consequently points A ($\omega/\omega_p=0.085$) and B ($\omega/\omega_p=0.2$) over-predicts and under-predicts the actual resonant frequencies of the two dipoles respectively (which are actually at $\omega/\omega_p=0.078$ and $0.234$).

It is worth mentioning that up to now we have restricted ourselves to two-dimensional cylindrical structures.  The Bohr model we have employed is very fundamental and the extension of this model to three-dimensional spherical structures is direct~\cite{Ruan2011_APL}. Specifically, Eq. (\ref{bohr condition})-Eq. (\ref{lsp_bohr condition}) can be directly applied: $k_{\rm spp}$ is the same for cylindrical and spherical structures as they share the same corresponding waveguide; also we can employ the same approximation for the effective phase accumulation path of cylindrical and spherical structures. As a simple demonstration that the Bohr model is also applicable to three-dimensional particles, in \rpict{fig3}(c) we show the extinction efficiency spectra of the dipole modes supported by the metal-dielectric-metal nanosphere [$R_{\rm eff}=(R_1+R_2)/2$].  The scattering configuration and the definition of the geometric parameters are the same as that of the nanowire (see \rpict{fig1} and \rpict{fig2}). As expected, two dipole modes are observed at different spectral positions, and the physical origin of the two modes is the same as that of the nanowire as we have discussed above.


\subsection*{Absorption maximization for localized surface plasmons}

From the discussions presented above, we know that the Bohr model offers us a new perspective for the understanding of LSP resonances based on SPP modes.  As a specific application of this model, we investigate the absorption maximization of LSP resonances based on those new understandings obtained.  The ideal case of loss maximization is the recent demonstration of coherent perfect absorption~\cite{Chong2010_PRL,Wan2011_science,Noh2012_PRL}. However, to achieve coherent perfect absorption for scattering particles, there are harsh restrictions on both the shape of the incident waves and the particles themselves~\cite{Chong2010_PRL,Wan2011_science,Noh2012_PRL}. It is hard to meet all those requirements simultaneously and thus more practical to discuss how to maximize the absorption of particles with plane wave incidence. This is certainly one of the most general problems, which might find applications in various fields and applications when the incident waves can be approximated as pane waves. 

\subsubsection*{Single-channel absorption limit}

\pict[0.99]{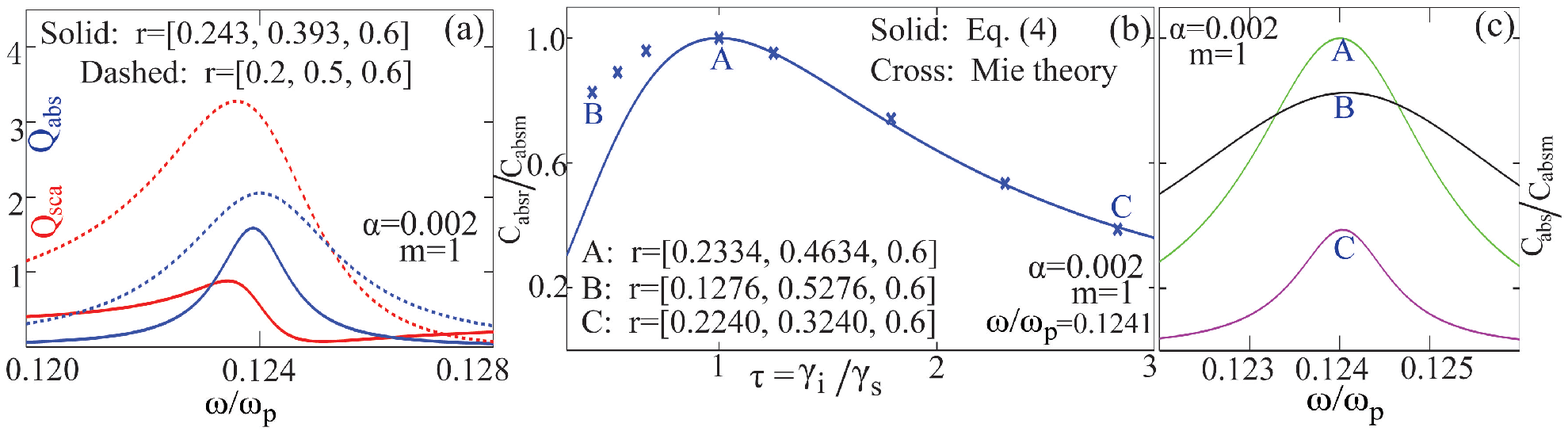}{fig4}{\small (a) Efficiencies of scattering cross sections (red curves) and absorption cross sections (blue curves) of the electric dipole modes ($m=1$) supported by two three-layered nanowires of different size parameters: $r=[0.2,~0.5,~0.6]$ (dashed curves) and $r=[0.243,~0.393,~0.6]$ (solid curves). The nanowire is the same as that shown in \rpict{fig2}(b) (inset), except that the refractive index of the dielectric layer has been changed to $n=2.1$. (b) Change of absorption cross section (normalized by the single channel absorption limit) of the dipole modes ($m=1$) with different loss ratios $\tau$.  The frequency is fixed at the resonant frequency of $\omega/\omega_p=0.1241$ as shown in (a).  The solid curve comes from Eq. (\ref{absorption_resonant}) and the crosses correspond to the calculated results of the three-layered nanowires (through Mie theory) with different size parameters. Three points A-C are pinpointed with the corresponding size parameters specified. The normalized cross section spectral at those three points are shown in (c).}

We still restrict our discussions to the simple structure of multi-layered nanowires.  For the two-dimensional cylindrical structures with plane wave incidence, the absorption cross section of a single resonance channel can be expressed as~\cite{Hamam2007_PRA,Ruan2010_PRL}:
\begin{equation}
\label{absorption}
{C_{\rm abs}} = {{2\lambda } \over \pi }{{{\gamma _i}{\gamma _s}} \over {{{(\omega  - {\omega _0})}^2} + {{({\gamma_i} + {\gamma _s})}^2}}},
\end{equation}
where $\lambda$ and $\omega$ are the wavelength and frequency of the incident wave respectively; $\omega_0$ is the resonant frequency; $\gamma_i$ and $\gamma_s$ are the intrinsic loss rate (due to Ohmic loss) and scattering loss rate (due to free-space scattering) respectively. At the resonant frequency, the expression can be simplified as;

\begin{equation}
\label{absorption_resonant}
 {C_{\rm absr}} = {{2\lambda } \over \pi({\tau  + {\tau ^{ - 1}}+2) }},
\end{equation}
 where $\tau=\gamma_i/\gamma_s$ is the loss ratio. Obviously the absorption cross sections can be maximized as ${C_{\rm absm}} = {\lambda  \over {2\pi }}$ when $\tau=1$, which requires that the  intrinsic loss and scattering loss rates are equal~\cite{Seok2011_radiation,Fleury2014_PRB,Tretyakov2013_arXiv,Estakhri2014_PRB}. ${C_{\rm absm}}$ is basically the single channel absorption limit.  However, when the wavelength of the incident wave is fixed, to tune the loss ratio to $\tau=1$ in order to reach the absorption limit without shifting the resonant wavelength is usually very challenging, which could involve complicated material engineering such as doping to change the intrinsic loss rate.

\subsubsection*{Absorption maximization through pure geometric tuning}

According to the geometric picture of the LSP resonances we have discussed, a fixed resonant frequency only requires that an integral number of $2\pi$ phase has been accumulated at this frequency, which can be satisfied for different phase accumulation paths with different dispersions of the SPP modes. Consequently the loss ratio could still be tuned through pure geometric tuning, without even shifting the resonant frequency.  We note here that in this paper our discussions have been restricted to loss maximization of an individual resonance channel, and the losses can be further enhanced through overlapping resonances of different orders~\cite{Hasegawa2006_OL,Estakhri2014_PRB}.

We employ again the three-layered nanowire shown in the inset of \rpict{fig2}(b). This time we substitute the $n=3.5$ dielectric layer with the $n=2.1$ dielectric layer. To decrease the refractive index of the dielectric layer will make smaller the effective wave-vector mismatch inside and outside the three-layered nanowire, and thus can increase the scattering loss rate $\gamma_s$~\cite{Yang2012_NP450},  rendering it comparable to $\gamma_i$. To make the two loss rates equal to each other requires extra geometric tuning of the structure. Instead of calculating separately $\gamma_s$ and $\gamma_i$ to decide the loss ratio of $\tau$, we calculate firstly the scattering and absorption cross sections of the nanowire through Mie theory and then the loss ratio can be obtained directly as the ratio of the absorption cross section to the scattering cross section~\cite{Hamam2007_PRA,Ruan2010_PRL}:
\begin{equation}
\label{loss_ratio}
\tau=\gamma_i/\gamma_s=C_{\rm abs}/C_{\rm sca}.
\end{equation}
 This is because that similar to Eq. (\ref{absorption}), the scattering cross section can be expressed as: ${C_{\rm sca}} = {{2\lambda } \over \pi }{{\gamma _s^2} \mathord{\left/
 {\vphantom {{\gamma _s \gamma _i} {[{{(\omega  - {\omega _0})}^2} + {{({\gamma _i} + {\gamma _s})}^2}]}}} \right.
 \kern-\nulldelimiterspace} {[{{(\omega  - {\omega _0})}^2} + {{({\gamma _i} + {\gamma _s})}^2}]}}$~\cite{Hamam2007_PRA,Ruan2010_PRL}.  \rpict{fig4}(a) shows both the scattering  efficiency (red curves) and absorption efficiency (blue curves) of the dipole modes ($m=1$) supported by two three-layered nanowires of different size parameters: $r=[0.2,~0.5,~0.6]$ (dashed curves) and $r=[0.243,~0.393,~0.6]$ (solid curves). When the size parameters are tuned, it is clear from \rpict{fig4}(a) that: (1) the resonance position of the absorption can still be fixed ($\omega/\omega_p=0.1241$) and (ii) as the ratio of the absorption cross section to the scattering cross section, the loss ratio $\tau$ can be effectively tuned.   \rpict{fig4}(b) shows the change of absorption cross section (normalized by the single channel absorption limit at $\omega/\omega_p=0.1241$) of the electric dipole modes ($m=1$) at the resonant frequency of $\omega/\omega_p=0.1241$ with different loss ratios $\tau$. The solid curve corresponds to  Eq. (\ref{absorption_resonant}) and the crosses correspond to the calculated results of nanowires (through Mie theory) with different size parameters. Three points A-C are pinpointed with the corresponding size parameters specified. Further details about absorption cross section spectra at those three points are shown in \rpict{fig4}(c). It is clear that the absorption can be effectively maximized through pure geometric tuning, which reaches the single channel absorption limit at the size parameter of $r=[0.2334,~0.4634,~0.6]$. We note here that there are some discrepancies between the results from Eq. (\ref{absorption_resonant}) and those from Mie theory [\rpict{fig4}(a)]. This is due to that the central resonant positions of the absorption and scattering spectra do not fully overlap  [\rpict{fig4}(a)], and consequently the relation in Eq. (\ref{loss_ratio}) is not exactly rigorous. Nevertheless this approximation does not at all obscure the physical mechanism and the evolution trend of absorption with changing loss ratios has been clearly revealed, as shown in \rpict{fig4}(b).

\section*{Discussion}

In this work, based on the Bohr model, we give a geometric picture for the LSP resonances supported by plasmonic nanoparticles relying on the SPP modes propagating in the corresponding plasmonic waveguides. This geometric model actually establishes a connection between the two sub-fields of plasmonics, and can be firstly applied to explain directly the well known scattering features of plasmonic nanoparticles, such as well defined LSP resonance in the subwavelength regime,  higher (lower) order modes supported at higher (lower) frequencies and redshift (blueshift) of the resonances with increasing (decreasing) particle sizes. As a next step, through this geometric model, it is further demonstrated that other anomalous scattering features can be obtained within geometrically engineered multi-layered plasmonic nanoparticles, including higher (lower) order modes supported at lower (higher) frequencies and blueshift (redshift) of the resonances with increasing (decreasing) particle sizes. We reveal that those anomalous scattering features originate from the existence of backward SPP modes in the corresponding waveguides. At the same time, it is shown that the multi-layered nanowires can be engineered to support multiple electric resonances of the same order at different frequencies with different near-field distributions. This unusual feature is induced by the multiple dispersion bands of the corresponding plasmonic waveguides. The number of the modes of the same order is decided by the number of the dispersion bands. At the end we show that at a fixed resonant frequency, the absorption of a plasmonic resonance can be effectively maximized through pure geometric tuning to reach the single channel absorption limit, when the scattering and absorption rates are made equal.

 The intuitive geometric picture should not be confined to plasmonic particles and can be actually applied to many other localized resonances in  other structures. Despite the simplicity and generality of this model, however there do exist some challenges with it: (i) As is shown in the discussions associated with \rpict{fig3}, it is not so simple or direct to rigorously decide the effective length of the phase accumulation path. This challenge becomes even tougher for structures with sharp corners. There have been some related attempts~\cite{Roll2000_JOSAA,Papoff2011_OE}, which however have neither the simplify nor clarity. Consequently they show no superiority to the Mie theory or direct simulations, thus offering very limited physical insights. (ii) A lot of work has to be done to make this model applicable and compatible with structures which incorporate materials with nonlocal, nonlinear and/or  quantum effects~\cite{Boardman1982_book,Scholl2012_nature,Kauranen2012_NP,Esteban2012_NC}. (iii) Our geometric model applies only to electric plasmonic resonances, and all the resonances discussed in this paper are electric resonances.  The geometric model for the optically-induced magnetic resonances~\cite{Kuznetsov2012_SciRep,Liu2014_CPB} is still not available. (iv) The Bohr model interpretation would lead to the conclusion that for all plasmonic particles with sizes getting smaller and smaller into the quasi-static spectral regime, the resonance frequency of all electric resonances would converge to the surface plasmon frequency (${{\omega _p } \over {\sqrt 2 }}$). This is exactly the case for nanowire scattering configuration we have investigated in our paper, but not the case for other three-dimensional structures (\textit{e.g.}, for spherical plasmonic nanoparticles, the electric dipole resonant frequency would converge to ${{\omega _p } \over {\sqrt 3 }}$ and for electric resonances of other orders it would converge to higher frequencies). But it is worth noticing that the Bohr model works well for spherical structures with particle sizes not far smaller than the wavelength [see \rpict{fig3}(c)].  At the end we would like to note here that in this paper we characterize resonances through far-field scattering spectra and we should keep in mind that resonances can be also described by near-field properties~\cite{Zuloaga2011_NL,Alonso2013_PRL}.

The intuitive Bohr model based geometric picture we discussed in this work offers new insights into the understanding of LSP resonances and other localized resonances. It might shed new light to investigations related to the particle scattering problem and moreover can possibly accelerate the merging of this field with other rapidly developing fields, such as  graphene~\cite{Geim2007_NM} and topological states of light and matter~\cite{Yin2013_science,Khanikaev2013_NM}. This could bring out a lot of extra opportunities, for not only fundamental studies, but also for a lot of particle scattering related applications, including biosensing, nanoantennas, photovoltaic devices, switching and even medical treatments.

\section*{Methods}

\subsection*{Dispersions of the three-layered metal-dielectric plasmonic waveguides}

The dispersion relation of the three-layered metal-dielectric plasmonic waveguide [shown in the inset of \rpict{fig2}(a)] can be obtained through solving the scalar \textit{Helmholtz} equation combined with the continuity equations for the fields at the boundaries. As we are interested only in the SPP modes and thus only the TM modes have been discussed. The dispersion relation for TM modes within this waveguide, which links the free-space frequency $\omega$ and the SPP wave-vector $k_{\rm spp}$, can be expressed as~\cite{Kudo1982_RS}:
\begin{equation}
\label{dispersion_4layer}
{e^{ - 4{\Delta _d}{\varepsilon _m}}} - {{t + s} \over {t - s}} = 0,
\end{equation}
where $s = {\Delta _d}{\Delta _m}({\Delta _m} + {\Delta _b})[1 + \tanh(2\delta {\Delta _m}{\varepsilon _m})]$; $t = {\Delta _m}(\Delta _d^2 + {\Delta _m}{\Delta _b}) + (\Delta _d^2{\Delta _b} + i{\Delta _d}\Delta _m^2)\tanh(2\delta {\Delta _m}{\varepsilon _m})$; $\tanh(\cdot)$ is the hyperbolic tangent function; $\delta  = {{{D_2}} \over {{D_1}}}$; ${\Delta _{m,d,b}} = {{{D_1}} \over 2}{\sqrt {k_{\rm spp}^2 - {\varepsilon _{m,d,b}}k_0^2}}$; $\varepsilon$ denotes the relative permittivity; ${k_0}$ is the wave-vector in vacuum; subscripts $m$, $d$, $b$ denote the metal layer, dielectric layer, and background media respectively. The dispersion curve can be obtained through solving the above equation.

\subsection*{Scattering of the plasmonic nanowires and nanospheres}

For the three-layered plasmonic nanowire (and nanosphere) shown in the inset of \rpict{fig2}(b), we obtain its scattering properties through applying the generalized Mie theory~\cite{Kerker1969_book}:

(1) For nanowires, the incident plane wave is polarized in plane (perpendicular to the nanowire axis). The expressions for the cross sections of extinction, scattering and absorption are~\cite{Kerker1969_book}:
\begin{eqnarray}
\label{expressions_scattering}
&{C_{\rm ext}} = {{2\lambda } \over \pi }{\mathop{\rm Re}\nolimits} \left[ {{a_0} + 2\sum\nolimits_{m = 1}^\infty  {{a_m}} } \right],\\
&{C_{\rm sca}} = {{2\lambda } \over \pi }\left[ {|a_0|^2 + 2\sum\nolimits_{m = 1}^\infty  {|a_m|^2} } \right],\\
&{C_{\rm abs}} = {C_{\rm ext}} - {C_{\rm sca}},
\end{eqnarray}
where $\rm Re[\cdot]$ means the real part; $\lambda$ is the wavelength in the background material (vacuum in this study); $a_0$ and $a_m$ are the scattering coefficients, whose expressions are long and tedious, which will not be shown here but can be found in Refs.~\cite{Kerker1969_book}. More specifically $a_0$ and $a_n~(n=1,~2,~...)$ correspond to the optically-induced magnetic dipole and electric modes of order $n$, respectively~\cite{Liu2013_OL2621}.

(2) For nanospheres, the cross sections are independent of the direction of the polarization:
\begin{eqnarray}
\label{expressions_scattering}
&{C_{\rm ext}} = {\lambda^2  \over {{2}{\pi}}}\sum\limits_{m = 1}^\infty  {(2m + 1){\mathop{\rm Re}\nolimits} ({a_m}}  + {b_m}),\\
&{C_{\rm sca}} = {\lambda^2  \over {{2}{\pi}}}\sum\limits_{m = 1}^\infty  {(2m + 1){\mathop{}\nolimits} ({|a_m|^2}}  + {|b_m|^2}),\\
&{C_{\rm abs}} = {C_{\rm ext}} - {C_{\rm sca}},
\end{eqnarray}
 where $a_m$ and $b_m$ are the scattering coefficients, whose expressions can be found in Refs.~\cite{Kerker1969_book}. More specifically $a_m$ and $b_m$ correspond to the  electric  and optically-induced magnetic  modes of order $m$, respectively~\cite{Liu2014_CPB,Liu2012_ACSNANO,Liu2012_PRB}.

It is obvious that the contributions from modes of specific orders can be calculated directly and separately. Our discussion in this paper are restricted to electric modes only.  The corresponding efficiencies are defined as:  the ratios of the cross sections to the diameter of the outmost layer of the nanowire (for nanowires) and the ratios of the cross sections to the area of the outmost layer of the nanosphere (for nanospheres).


\section*{Acknowledgements}

We are indebted to A. E. Miroshnichenko, D. N. Neshev, and J. Hou for many useful discussions and suggestions. This work has been supported by  the National Natural Science Foundation of China (Grant No.: $11404403$), the Australian Research Council through the Center of Excellence CUDOS, the Leverhulme Trust of UK, and the Basic Research Scheme of College of Optoelectronic Science and Engineering, National University of Defense Technology.

\section*{Author contributions statement}

W. L. conceived the idea and performed the calculations; W. L.,  R. F. O. and Y. S. K. discussed the
results and analysed the data; W. L. wrote the paper with a lot of input from  R. F. O. and Y. S. K..

\section*{Additional information}

\textbf{Competing financial interests:} The authors declare no competing financial interests.

%

\end{document}